\documentclass[showpacs,twocolumn,superscriptaddress,
preprintnumbers,amsmath,amssymb,prl,reprint]{revtex4}
\usepackage{graphicx}
\usepackage{bm}
\usepackage{color}
\usepackage{url}
\usepackage{datetime}

\begin{document}

\title{Improved Measurement of the Hydrogen  $\bm{1S}$\,--\,$\bm{2S}$ Transition Frequency}

\author{Christian G. Parthey}
    \altaffiliation{The first two authors contributed to this work equally.}
\affiliation{Max-Planck-Institut f\"{u}r Quantenoptik, 85748 Garching, Germany}
\author{Arthur Matveev}
    \altaffiliation{The first two authors contributed to this work equally.}
\affiliation{Max-Planck-Institut f\"{u}r Quantenoptik, 85748 Garching, Germany}
\author{Janis Alnis}
\affiliation{Max-Planck-Institut f\"{u}r Quantenoptik, 85748 Garching, Germany}
\author{Birgitta Bernhardt}
\affiliation{Max-Planck-Institut f\"{u}r Quantenoptik, 85748 Garching, Germany}
\author{Axel Beyer}
\affiliation{Max-Planck-Institut f\"{u}r Quantenoptik, 85748 Garching, Germany}
\author{Ronald Holzwarth}
	\altaffiliation{Also at Menlo Systems GmbH, Martinsried, Germany}
\affiliation{Max-Planck-Institut f\"{u}r Quantenoptik, 85748 Garching, Germany}
\author{Aliaksei Maistrou}
\affiliation{Max-Planck-Institut f\"{u}r Quantenoptik, 85748 Garching, Germany}
\author{Randolf Pohl}
\affiliation{Max-Planck-Institut f\"{u}r Quantenoptik, 85748 Garching, Germany}
\author{Katharina Predehl}
\affiliation{Max-Planck-Institut f\"{u}r Quantenoptik, 85748 Garching, Germany}
\author{Thomas Udem}
\affiliation{Max-Planck-Institut f\"{u}r Quantenoptik, 85748 Garching, Germany}
\author{Tobias Wilken}
\affiliation{Max-Planck-Institut f\"{u}r Quantenoptik, 85748 Garching, Germany}
\author{Nikolai Kolachevsky}
        \altaffiliation{Also at P.~N.~Lebedev Institute, Moscow}
    \email{kolik@lebedev.ru}
\affiliation{Max-Planck-Institut f\"{u}r Quantenoptik, 85748 Garching, Germany}
\author{Michel Abgrall}
\affiliation{LNE-SYRTE, Observatoire de Paris, 61 avenue de l'Observatoire, 75014 Paris, France}
\author{Daniele Rovera}
\affiliation{LNE-SYRTE, Observatoire de Paris, 61 avenue de l'Observatoire, 75014 Paris, France}
\author{Christophe Salomon}
\affiliation{Laboratoire Kastler-Brossel, CNRS, 24 rue Lhomond, 75231 Paris, France}
\author{Philippe Laurent}
\affiliation{LNE-SYRTE, Observatoire de Paris, 61 avenue de l'Observatoire, 75014 Paris, France}
\author{Theodor W. H\"ansch}
        \altaffiliation{Also at Ludwig-Maximilians-University, Munich}
        \affiliation{Max-Planck-Institut f\"{u}r Quantenoptik, 85748 Garching, Germany}

\begin{abstract}
We have measured the $1S-2S$ transition frequency in atomic hydrogen via two photon spectroscopy on a 5.8\,K atomic beam. We obtain $f_{1S-2S} = 2\,466\,061\,413\,187\,035\,(10)\,\text{Hz}$ for the hyperfine centroid. This is a fractional frequency uncertainty of $4.2\times10^{-15}$ improving the previous measurement by our own group [M.~Fischer {\em et al.}, Phys. Rev. Lett. {\bf 92}, 230802 (2004)] by a factor of 3.3. The probe laser frequency was phase coherently linked to the mobile cesium fountain clock FOM via a frequency comb.\par 
\end{abstract}

\maketitle

For the last five decades spectroscopy on atomic hydrogen along with its calculable atomic structure have been fueling the development and testing of quantum electro-dynamics (QED) and has lead to a precise determination of the Rydberg constant and the proton charge radius~\cite{Biraben09}. The absolute frequency of the $1S-2S$ transition has been measured with particularly high precision, so that it now serves as a corner stone in the least squares adjustment of the fundamental constants~\cite{CODATA06}. The resonance has been used to set limits on a possible variation of fundamental constants~\cite{Fischer04} and violation of Lorentz boost invariance~\cite{Altschul10}. It further promises a stringent test of the charge conjugation/parity/time reversal (CPT) theorem by comparison with the same transition in antihydrogen \cite{Andresen11}. \par

In this Letter we present a more than three times more accurate measurement of the $1S-2S$ transition as compared to the previous best measurements~\cite{Niering00,Fischer04} reaching the $4\times10^{-15}$ regime. The key improvements are the replacement of a dye laser by a diode laser system with improved frequency stability for the two photon spectroscopy. In addition, a direct measurement of the $2S$ velocity distribution of the thermal atomic hydrogen beam allows a more accurate characterization of the second order Doppler effect. Finally, a quench laser resetting the population to the ground state right after the hydrogen nozzle was introduced. This removes possible frequency shifts due to the high density of atoms and a possible dc Stark shift within the nozzle.\par

The extended-cavity diode spectroscopy laser system (ECDL) has been described elsewhere~\cite{Kolachevsky11}. An ECDL master oscillator near $972\,\text{nm}$ is amplified in a tapered amplifier. The laser radiation is frequency doubled twice within two resonant cavities to obtain 13\,mW of the required uv light near $243\,\text{nm}$. Locking the laser to a high finesse cavity made from ultra-low expansion glass leads to a line width of less than $1\,\text{Hz}$ and a fractional frequency drift of $1.6\times10^{-16}\,\text{s}^{-1}$~\cite{Alnis08}. A fiber laser frequency comb with a repetition rate of 250\,MHz is used to phase-coherently link the cavity frequency to an active hydrogen maser which is calibrated using the mobile cesium fountain atomic clock FOM~\cite{FOM}. We apply cycle slip detection as described in \cite{Udem98}.\par

\begin{figure}[t!]
    \begin{center}
    \includegraphics [width=\columnwidth]{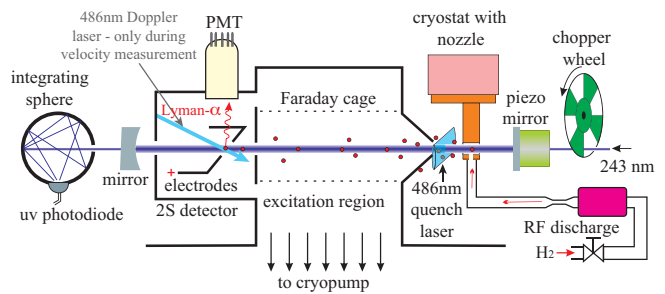}
    \caption{(color online). Schematic of the beam apparatus. A standing laser wave at 243\,nm (between grey mirrors) with a $1/e^2$ waist radius of $w_0 = 292\,\mu\text{m}$ at the flat cavity front mirror excites the sharp $1S-2S$ transition in a co-linearly propagating cold thermal beam of atomic hydrogen emerging     from a cooled copper nozzle. The $2S$ state is detected after quenching with a localized electric field which releases a Lyman-$\alpha$ photon. See text for further details.}\label{fig:setup}
    \end{center}
\end{figure}

The beam apparatus (Fig.~\ref{fig:setup}) follows the design of~\cite{Walraven82} and has been used before~\cite{Parthey10}. Hydrogen is dissociated at a pressure of 1\,mbar in a radio frequency (rf) discharge running in a sapphire tube. A Teflon capillary controls the flow and Teflon tubing guides the atomic hydrogen to a copper nozzle cooled to 5.8\,K by a liquid helium flow cryostat. The dissociation fraction at the nozzle (2.2\,mm diameter) is about 10\,\%. After 45\,min of operation the nozzle closes up with frozen molecular hydrogen. We then heat it to 20\,K for 15\,min to evaporate the hydrogen molecules. The atomic beam is defined by a 2.4\,mm (front) and a 2.1\,mm (rear) diameter aperture which also separate the differentially pumped excitation region ($10^{-5}\,\text{mbar}$ / $10^{-8}\,\text{mbar}$). This region is shielded by a grounded Faraday cage from electric stray fields that may build up from laser ionized hydrogen. An enhancement cavity with a finesse of 120 forms a standing wave for collinear excitation of the $1S-2S$ transition by two counter-propagating photons. For detection, excited, metastable atoms are quenched via the $2P$ state by an electric field (10\,V/cm) and the emitted 121\,nm photons are detected by a photon multiplier tube (PMT). The intra-cavity power is monitored by measuring the transmission of the cavity using a photo diode connected to an integrating sphere.\par

During twelve consecutive days starting on May 30 2010 with a break on June 9, 1587 $1S-2S$ spectra have been recorded. For each $1S-2S$ line the spectroscopy laser's frequency samples the transition in random order. At each frequency point we count Lyman-$\alpha$ photons for one second (with a 50\,\% duty cycle due to the chopper) at two different 243\,nm laser intensities. A double pass AOM in zeroth order placed in front of the enhancement cavity allows to quickly alter the power level under otherwise identical conditions. After switching the AOM we wait several milliseconds to avoid possible frequency chirps.\par

We apply an external magnetic field of $0.5\,\text{mT}$ to separate the hyperfine components. For the spectroscopy, we use the transitions from $F=1, m_F=\pm1$ to $F'=1, m_{F'}=\pm1$ whose Zeeman shifts cancel to zero. The hyperfine centroid frequency can be obtained by adding $\Delta f_{\text{HFS}} = +\,310\,712\,229.4\,(1.7)\,\text{Hz}$ calculated from the experimental results for the $1S$ and $2S$ hyperfine splittings~\cite{Essen71,Kolachevsky09}.\par

Now, we discuss the compensation of the two main systematic effects and estimations of the remainder. The Doppler effect, due to the velocity $v$ of the atoms is cancelled to first order by virtue of the two photon excitation scheme~\cite{Haensch75}. The remaining second order Doppler shift $\Delta f_{\text{dp}}=-v^2f_{1S-2S}/(2c^2)$ is compensated in two steps: First, we chop the excitation light at 160\,Hz (see Fig.~\ref{fig:setup}) which allows time of flight resolved detection. After the light is switched off we only start recording the $2S$ signal after certain delays~$\tau$ to let the fastest atoms escape. This samples the slow tail of the velocity distribution. The $2S$ counts are then sorted into twelve time bins $\tau_1=10\dots210\,\mu\text{s}$, $\tau_2=210\dots410\,\mu\text{s}$, $\dots$, $\tau_{12}=2210\dots2410\,\mu\text{s}$. This removes most of the second order Doppler effect, but is not sufficient to meet the current level of statistical uncertainty. Therefore, we measure the velocity distribution of the $2S$ atoms at the detection point as described below and use it to further correct the second order Doppler effect. The residual uncertainty of this procedure was determined by evaluating a second data set that was generated using a Monte Carlo (MC) simulation~\cite{Kolachevsky06,Haas06} in exactly the same way. For various simulation parameters such as temperature, geometry and initial $1S$ velocity distributions we find the uncertainty to be smaller than $2.0\times10^{-15}$ which is below the current statistical uncertainty of $2.6\times10^{-15}$ (see Tab.~\ref{tab:uncertainties}).\par

\begin{figure}[t!]
    \begin{center}
    \includegraphics [width=\columnwidth]{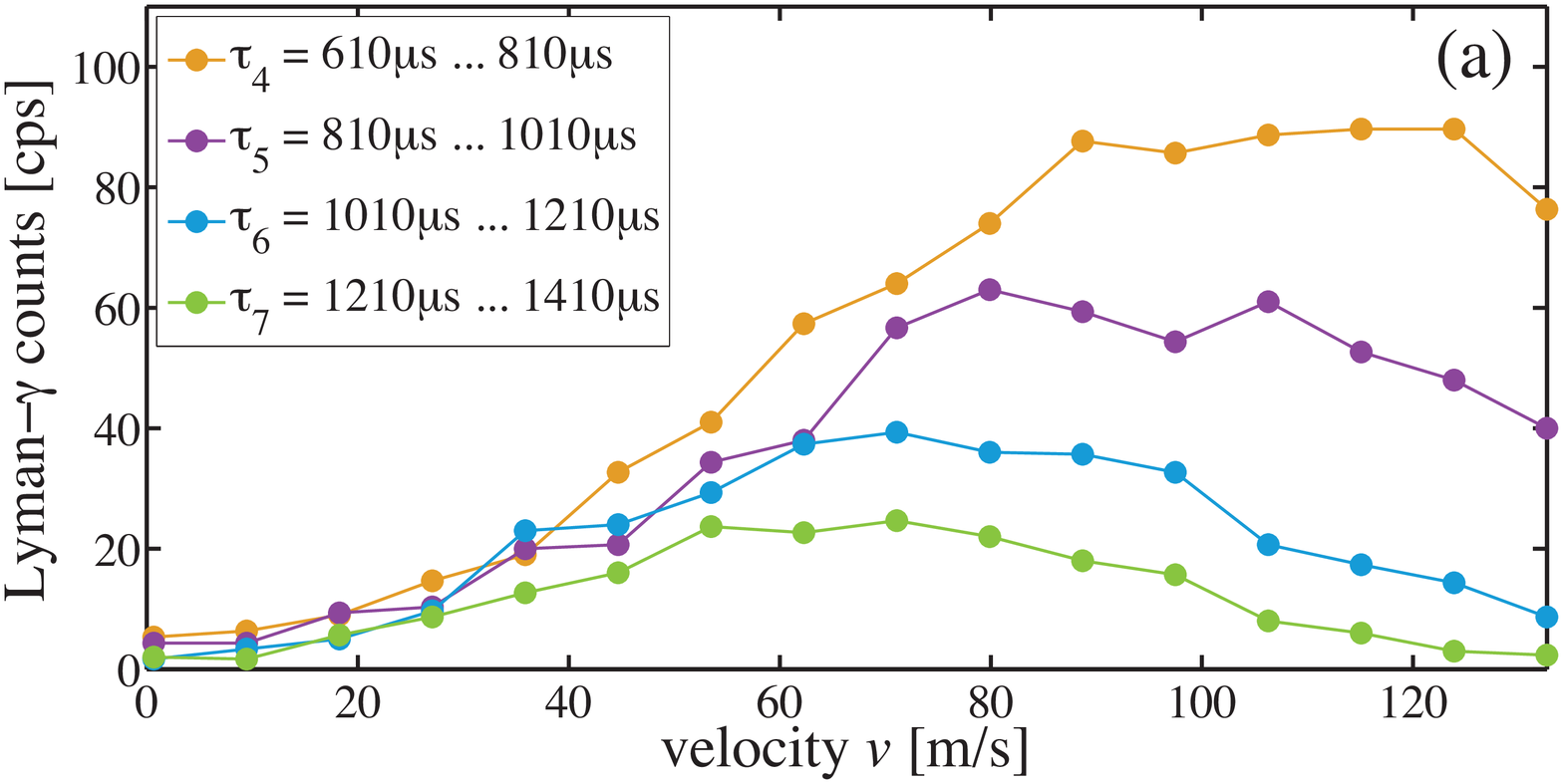}\\
    \includegraphics [width=\columnwidth]{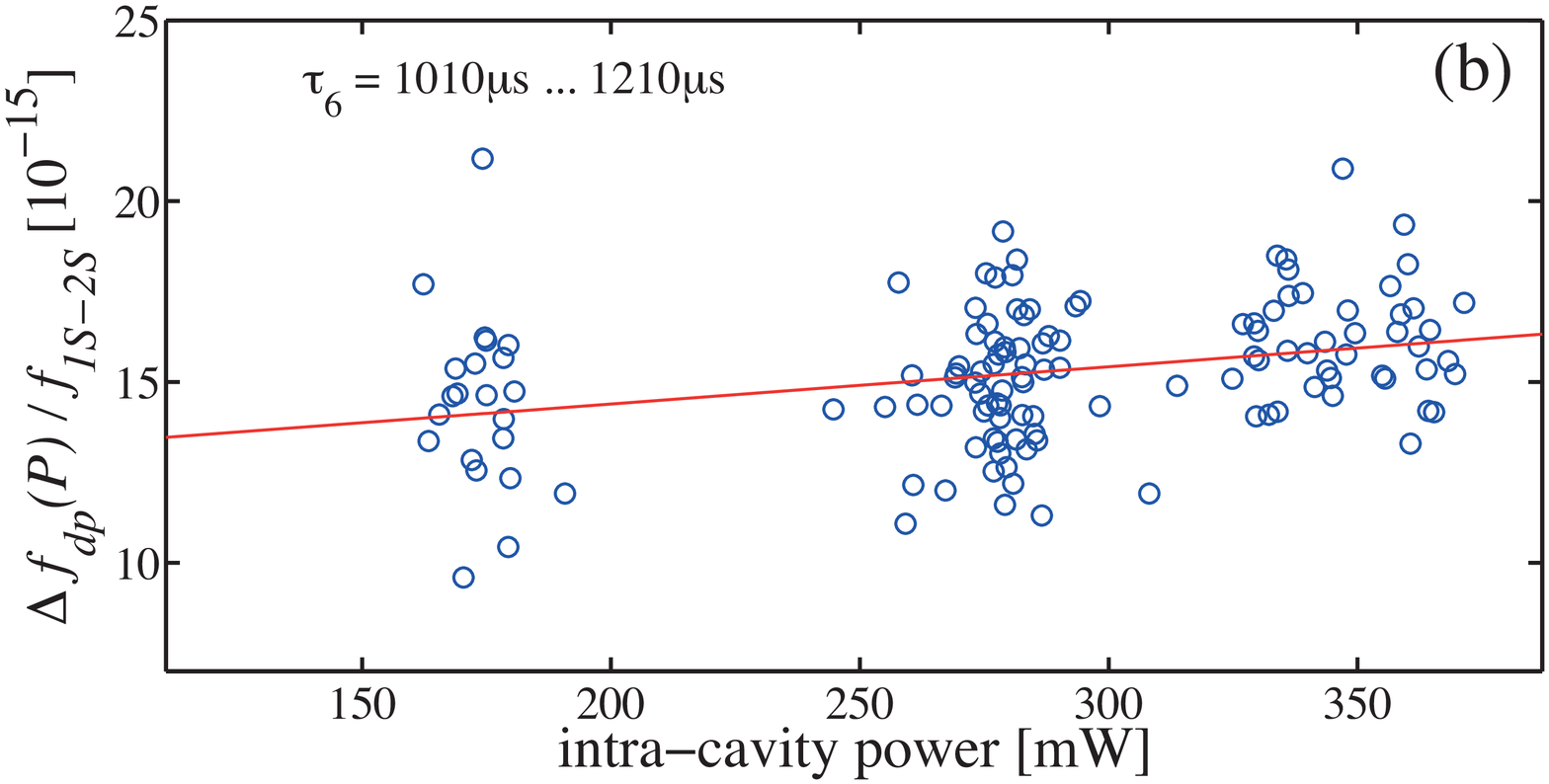}
    \caption{(color online). (a) Low velocity part of the Doppler profiles of $2S$ atoms recorded for different delays $\tau$ at $370\,\text{mW}$ intra-cavity power. (b) Fractional second order Doppler correction $\Delta f_{\text{dp}}(P)/f_{1S-2S}$ versus intra-cavity power for delay $\tau_{6}=1010\,\mu\text{s}\dots1210\,\mu\text{s}$. Each point represents the Doppler correction     as calculated from the central velocity of a single velocity profile measurement along with a     linear fit.}\label{fig:doppler_profile}
    \end{center}
\end{figure}

The second systematic effect that needs to be corrected for is the ac Stark shift which is mostly linear in laser power $P$. However, a small quadratic contribution~\cite{Kolachevsky06} must be taken into account, before we can apply a linear extrapolation using the stable, but otherwise not precisely calibrated laser power readings. The main contribution to this non-linear ac Stark shift is due to ionization of the $2S$ atoms by a third 243\,nm photon that removes preferably atoms that see larger laser powers. For excitation laser powers of 300\,mW as present in our experiment the quadratic ac Stark shift contributes on the order of $1\times10^{-14}$ as derived from the MC simulations. This is sufficiently small to rely on these simulations that assume a Maxwell distribution for the $1S$ atoms using the absolute laser power within 20\,\% relative uncertainty. Modeling and subtracting the delay dependent quadratic ac Stark effect in this way then allows to linearly extrapolate the line centers, without knowing the exact laser power calibration. This procedure reduces the overall ac Stark shift uncertainty to $0.8\times10^{-15}$.\par

We find the experimental (and simulated) line centers by fitting Lorentzians which represents a good approximation of the line shape for delays $\tau_4=610\dots810\,\mu\text{s}$ and higher. For lower delays the second order Doppler effect causes an asymmetry so we do not evaluate them. A small residual asymmetry for the longer delays is determined and taken into account by comparing with the MC simulation. Again, we use this simulation only for small corrections.\par

To correct the second order Doppler effect $\Delta f_{\text{dp}}=-v^2f_{1S-2S}/(2c^2)$, an accurate understanding of the velocity distribution is desirable. Previously, this information has been extracted from the line shape of the $2S$ spectra with an uncertainty of $8\times10^{-15}$~\cite{Niering00}. Here, we measure the $2S$ velocity distribution directly via the first order Doppler effect on the $2S-4P$ one photon transition which we excite under an angle of $45^{\circ}$ near 486\,nm (Doppler laser in Fig.~\ref{fig:setup}). The $2S-4P$ transition has a sufficiently narrow natural line width of 13\,MHz (corresponding to $\Delta v=8\,\text{m/s}$ at $45^{\circ}$) to resolve velocities on the level of 1\,m/s.\par

The $4P$ state decays to the ground state with a 90\,\% branching ratio emitting a 97\,nm Lyman-$\gamma$ photon which can be easily detected using a channeltron. Pulsing the 486\,nm Doppler laser with an AOM avoids power broadening (and loss of velocity resolution) while providing equal quench probability for atoms of different velocity. We cross the 486\,nm beam with the atomic beam right between the quench electrodes which are grounded for the velocity distribution measurements. Using the same delayed detection as for the $1S-2S$ spectra allows to extract the velocity distribution of $2S$ atoms that contribute to the signal with delay $\tau$. Working with the excited state directly gives the advantage of measuring the convolution of the velocity distribution with the excitation probability making the simulation of excitation dynamics unnecessary for this purpose. The low velocity part of a typical Doppler profile for delays $\tau_4=610\dots810\,\mu\text{s}$ to $\tau_7=1210\dots1410\,\mu\text{s}$ is shown in Fig.~\ref{fig:doppler_profile}~(a).\par

\begin{figure}[tb!]
    \begin{center}
    \includegraphics [width=\columnwidth]{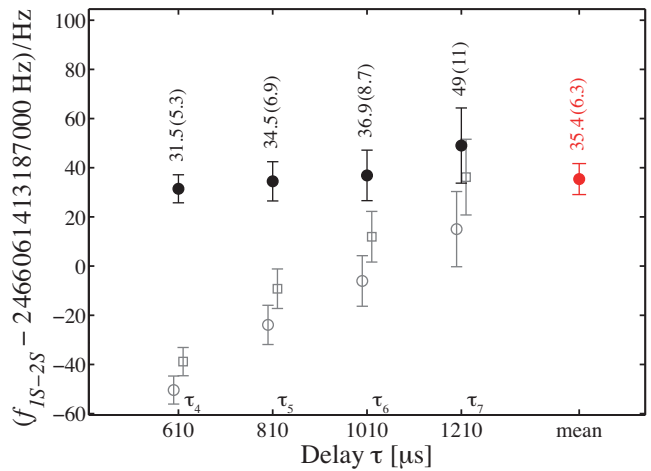}
    \caption{(color online). The plot shows $f_{1S-2S}$ with all corrections applied for     delays $\tau_4=610\dots810\,\mu\text{s}$ to $\tau_7=1210\dots1410\,\mu\text{s}$ (filled points) along with the mean as calculated from delays $\tau_{5,6}$ (see text). The error bars     represent the $1\sigma$ statistical uncertainty. The open circles show the same data     with the second order Doppler effect not corrected, the open squares do neither include the second oder Doppler effect nor the quadratic ac Stark shift corrections.}\label{fig:result_corrections}
    \end{center}
\end{figure}

From 131 recorded Doppler profiles $p_{\tau}(v,P)$ we calculate the second order Doppler effect for each delay~$\tau$ according to $\Delta f_{\text{dp}}(P) = -v_\text{c}^2(P)f_{1S-2S}/(2c^2)$ where the central velocity $v_\text{c}$ is determined by a Gaussian fit to the velocity profiles (see Fig.~\ref{fig:doppler_profile}\,(a)). The power dependence arises from ionization losses of slow atoms. A linear fit to $\Delta f_{\text{dp}}(P)$ reveals the second order Doppler effect correction as shown in Fig.~\ref{fig:doppler_profile}\,(b).\par

The uncertainty in the second order Doppler correction is caused by three main sources. First, the statistical uncertainty obtained from linear regression analysis of $\Delta f_{\text{dp}}(P)$ contributes $1.7\times10^{-15}$. Second, during the velocity measurements the $1S-2S$ spectroscopy laser was kept on the resonance only within $\pm160\,\text{Hz}$. MC simulation reveals an associated uncertainty of  $0.8\times10^{-15}$. Third, the $45^{\circ}$ angle between the atomic beam and the laser beam used to measure the velocity distribution can only be adjusted within $\pm1^{\circ}$. This translates to an uncertainty in the second order Doppler effect of $0.8\times10^{-15}$. Summing in quadrature leads to an overall uncertainty of the second order Doppler correction of $2.0\times10^{-15}$.\par


The fully corrected data is shown in Fig.~\ref{fig:result_corrections}. The transition frequency $f_{1S-2S}$ is independent of the delay for $\tau_4=610\dots810\,\mu\text{s}$ to $\tau_7=1210\dots1410\,\mu\text{s}$ unlike before the correction. The insignificant remaining slope can be readily explained within the uncertainty in the correction of the second order Doppler effect. For the final analysis we only use delays $\tau_{5,6}$. For delays $\tau_{1\dots4}$ the second order Doppler effect cannot be sufficiently characterized. For delay $\tau_{7}$ and higher, the quadratic ac Stark shift correction cannot be extracted with competitive uncertainty due to the inaccuracy of the absolute power measurement. Also, the statistics is poor for these delays.\par

\begin{table}[b]
\caption{Uncertainty budget. $\sigma$ - uncertainty.}\label{tab:uncertainties}
\begin{center}
\begin{ruledtabular}
    \begin{tabular}{l|c|c}
     & $\sigma$ & $\sigma/f_{1S-2S}$\\
     &  [Hz]    &    [$10^{-15}$] \\ \hline
    statistics      &   6.3  &  2.6 \\
    2nd order Doppler effect     &  5.1     &   2.0 \\
    line shape model    & 5.0   & 2.0 \\
    quadratic ac Stark shift (243\,nm)    &  2.0  &  0.8 \\
    ac Stark shift, 486\,nm quench light     &  2.0 &   0.8 \\
    hyperfine correction         &  1.7 &       0.69    \\
    dc Stark effect &   1.0 &   0.4 \\
    ac Stark shift, 486\,nm scattered light &   1.0 &   0.4 \\
    Zeeman shift & 0.93 & 0.38 \\
    pressure shift  &   0.5 &   0.2 \\
    blackbody radiation shift   &   0.3 &   0.12    \\
    power modulation AOM chirp  &   0.3 &   0.11    \\
    rf discharge ac Stark shift &   0.03    &   0.012   \\
    higher order modes  &   0.03    &   0.012   \\
    line pulling by $m_F=0$ component   &   0.004    &      0.0016\\
    recoil shift   &    0.009    &      0.0036 \\
    FOM &   2.0 &       0.81    \\
    gravitational red shift &   0.04    &       0.077   \\  \hline
    total   &   10.4     &  4.2 \\
    \end{tabular}
\end{ruledtabular}
\end{center}
\end{table}%

As mentioned above, the pressure shift and the dc Stark effect have been greatly reduced as compared to~\cite{Parthey10} by the introduction of the quench laser. With an atom flux of $10^{17}$ particles per second we find the remaining pressure shift within the excitation region to be well below $0.2\times10^{-15}$~\cite{McIntyre88}. In~\cite{Jentschura11} the remaining dc Stark shift was measured to be $0.4\times10^{-15}$.\par

\begin{figure}[t]
    \begin{center}
    \includegraphics [width=\columnwidth]{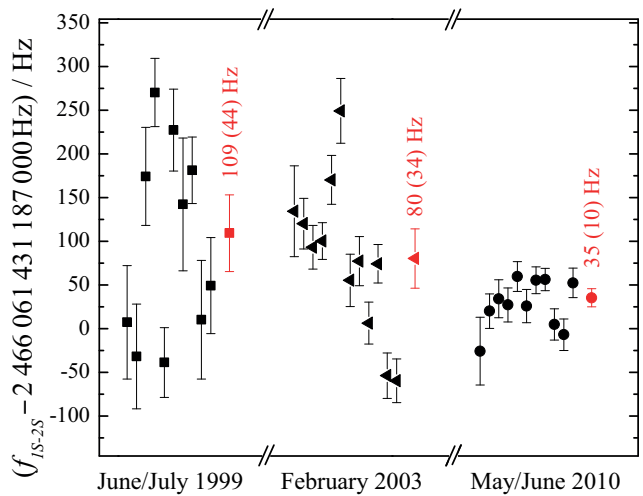}
    \caption{(color online). The $1S-2S$ centroid frequency of the current measurement is compared with the previous two measurements~\cite{Niering00,Fischer04}. The black points represent an average per day, the red points (with label) are the final value with the error bar representing the $1\sigma$ uncertainty.}\label{fig:comparison}
    \end{center}
\end{figure}

Scattered light from the 486\,nm quench beam can cause an ac Stark shift on the $1S-2S$ transition. We experimentally restrict such an effect on the $0.4\times10^{-15}$ level by intentionally increasing the stray light. Also, atoms leaving the quench beam experience a low intensity region at the far wing of the Gaussian beam profile at which they are shifted but not quenched. We numerically simulate this effect by extending our MC simulation by a spatially dependent quench rate and an ac Stark shift due to the quench laser. We find no significant shift with an uncertainty of $0.8\times10^{-15}$.\par

The Zeeman shift of the $F=1, m_F=\pm1$ to $F'=1, m_{F'}=\pm1$ hyperfine transitions is $\pm360\,\text{Hz/mT}$ and consequently averages to zero for equal populations in both $m_F$ components. In a dedicated experiment, we have measured the Zeeman shift to be below $0.38\times10^{-15}$ by varying the applied magnetic field.

Following~\cite{Farley81}, we correct the shift due to blackbody radiation at 300(30)\,K as $-4.1(1.2)\times10^{-16}$. Measuring the temperature fluctuations of the power switching AOM with the fed rf power we find an upper limit for an associated frequency chirp at $1.1\times10^{-16}$. From a measurement of the power radiated from the rf discharge into the excitation region we can estimate a possible ac Stark effect below $1.2\times10^{-17}$. Theoretical considerations limit the shift due to higher order modes in the enhancement cavity to $1.2\times10^{-17}$ and that by line pulling by the $m_F=0$ hyperfine component to $1.6\times10^{-18}$. A residual recoil shift can be restricted by analysis of the observed Doppler broadening to be below $3.6\times10^{-18}$.

The uncertainty of the fountain clock has been evaluated to $0.81\times10^{-15}$~\cite{FOMeval}. A conservative estimate of the height difference between clock and experiment contributes a gravitational red shift uncertainty of $7.7\times10^{-17}$. The uncertainty budget is summarized in Tab.~\ref{tab:uncertainties}.\par

Summarizing all corrections and uncertainties we find the $1S-2S$ {\em hyperfine centroid} frequency to be $f_{1S-2S} = 2\,466\,061\,413\,187\,035\,(10)\,\text{Hz}$. This corresponds to a fractional frequency uncertainty of $4.2\times10^{-15}$ and is in good agreement with our previous best measurement~\cite{Fischer04} but 3.3 times more accurate. A comparison of the previous two measurements along with the current result is presented in Fig.~\ref{fig:comparison}. The excessive day to day scatter present in the 1999 and 2003 measurements has been attributed to a dye laser instability and has consequently been removed in the current measurement.\par

The authors thank E.A.\,Hessels for insightful discussions, M.\,Fischer for providing a second frequency comb, and T.\,Nebel and M.\,Herrmann for carefully reading the manuscript. J.A.~acknowledges support by the Marie Curie Intra European program, N.K.~ from Presidential Grant MD-669.2011.8, and T.W.H.~from the Max Planck Foundation.

\end{document}